\documentclass[twoside,a4paper,11pt]{sea10}
\usepackage{graphicx}
\usepackage{hyperref}
\begin{document}
\pagenumbering{arabic}
\pagestyle{myheadings}
\thispagestyle{empty}
{\flushleft\includegraphics[width=\textwidth,bb=58 650 590 680]{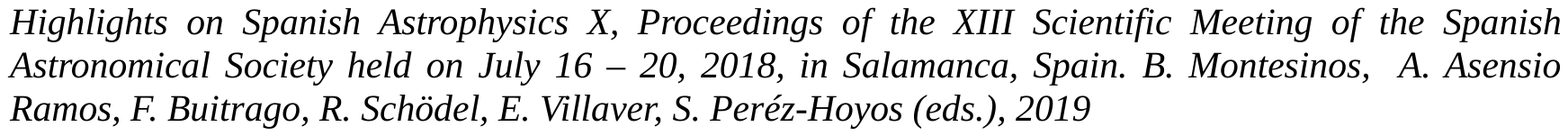}}
\vspace*{0.2cm}
\begin{flushleft}
{\bf {\LARGE
%
The binary central stars of planetary nebulae
%
}\\
\vspace*{1cm}
%
David Jones$^{1,2}$
%
}\\
\vspace*{0.5cm}
%
$^{1}$
Instituto de Astrof\'isica de Canarias, E-38205 La Laguna, Tenerife, Spain\\
$^{2}$
Departamento de Astrof\'isica, Universidad de La Laguna, E-38206 La Laguna, Tenerife, Spain
%
\end{flushleft}
%
\markboth{
The binary central stars of planetary nebulae
}{ 
%
David Jones
%
}
\thispagestyle{empty}
\vspace*{0.4cm}
\begin{minipage}[l]{0.09\textwidth}
\ 
\end{minipage}
\begin{minipage}[r]{0.9\textwidth}
\vspace{1cm}
\section*{Abstract}{\small
%
It is now clear that central star binarity plays a key role in the formation and evolution of planetary nebulae, with a significant fraction playing host to close-binary central stars which have survived one or more common envelope episodes.  Recent studies of these systems have revealed many surprises which place important constraints on the common envelope - a critical phase in the formation of a wide variety of astrophysical phenomena, including the cosmologically important supernovae type \textsc{i}a and other transient phenomena which will be detected by next-generation facilities, like the Large Synoptic Survey Telescope and the space-based gravitational wave detector the Laser Interferometer Space Antenna.

%
\normalsize}
\end{minipage}
%
%
%
\section{Introduction \label{intro}}

It is now clear that binary interactions play a key role in the shaping of planetary nebulae (PNe; \cite{hillwig16}) and that the binary fraction among PNe must be at least 20\% \cite{miszalski09a} but perhaps as high as 80\% \cite{demarco15,douchin15}.  However, in spite of their importance in understanding the origins of PNe, only some $\sim$60 binary central stars are known\footnote{A complete list is maintained at \href{http://drdjones.net/bCSPN}{http://drdjones.net/bCSPN}.} and, moreover, very little is known about these systems beyond their orbital periods.  In an attempt to reach a statistically significant sample of binary central stars, the author and collaborators are undertaking an observational campaign to search for and characterise these systems.  The search for new binary central stars has focussed primarily on targeted photometric and spectroscopic monitoring of the central stars of PNe which present morphological features believed to be characteristic of a binary evolution \cite{miszalski09b}.  These searches have been extremely successful leading to a rapid increase in the rate of discovery \cite{corradi11,boffin12,jones14,jones15,sowicka17,jones18}.  Detailed studies of these systems and those from the literature have revealed many interesting results with important implications for our understanding of close-binary evolution and the common envelope (CE) phase, in particular.

\begin{figure}
\center
\includegraphics[width=\textwidth,trim={0 2.5cm 0 2.5cm},clip=true]{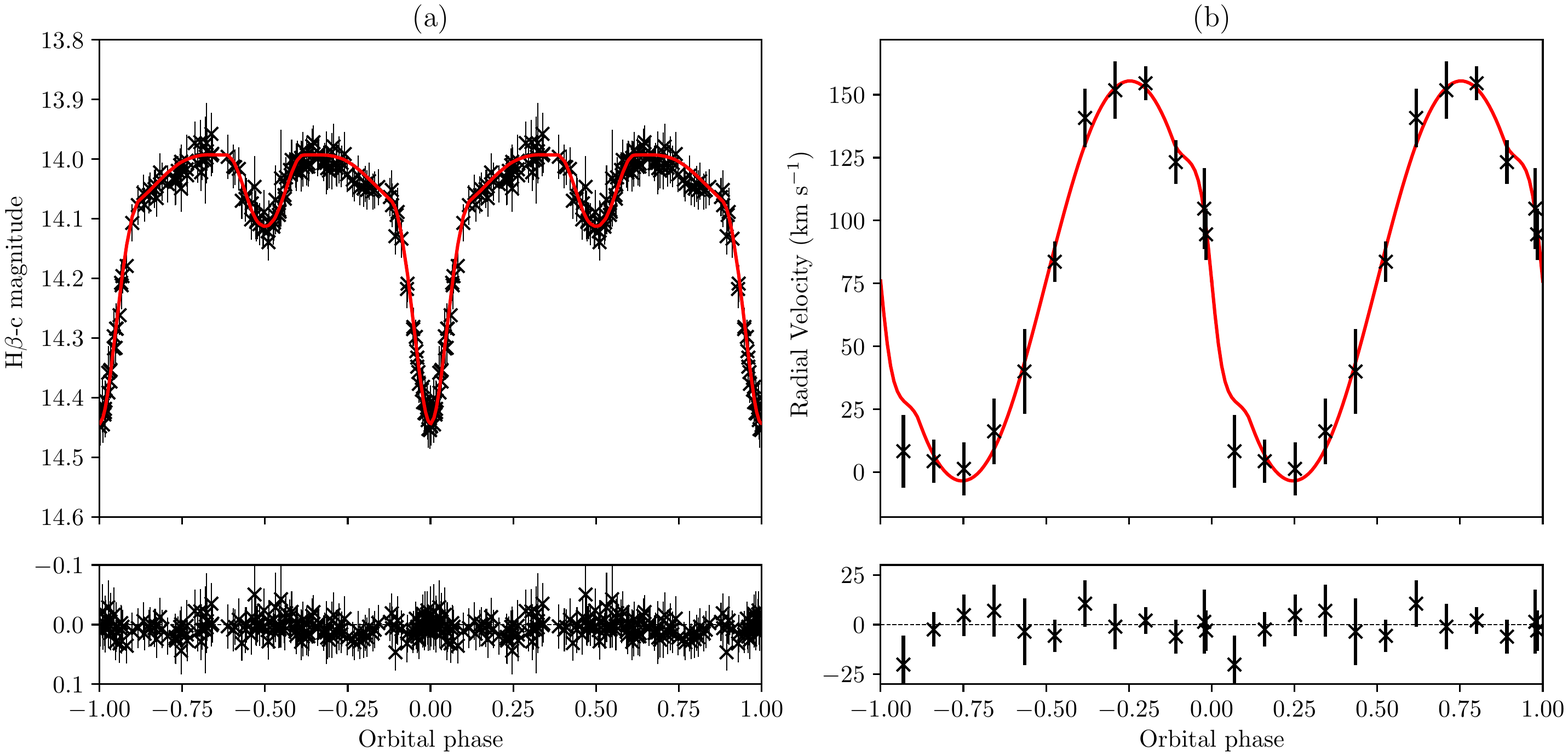} 
\caption{\label{djonesfig1} An example of the simultaneous modelling of light (a) and radial velocity (b) curves required to derive stellar (mass, temperature, radii) and binary (period, orbital inclination) parameters, in this case the central star of M~3-1 (reproduced from \cite{jones18}).
}
\end{figure}

\section{Probes of common envelope evolution}

The close-binary central stars of PNe are thought to have formed via a CE interaction, where the nebula itself is the remnant ejected envelope (see \cite{jones17a} for a review).  The presence of the short-lived ($\tau\leq$ 30,000 years) nebula makes these systems of particular interest for studying the CE, as they offer a unique opportunity to directly observe the properties of the ejected envelope (see, for example, Santander-Garc\'ia in these proceedings) as well as observe the stars themselves ``fresh-out-of-the-oven'', before the system has had time to adjust.

\begin{figure}
\center
\includegraphics[width=0.5\textwidth,angle=270,trim={0 0cm 0 0cm},clip=true]{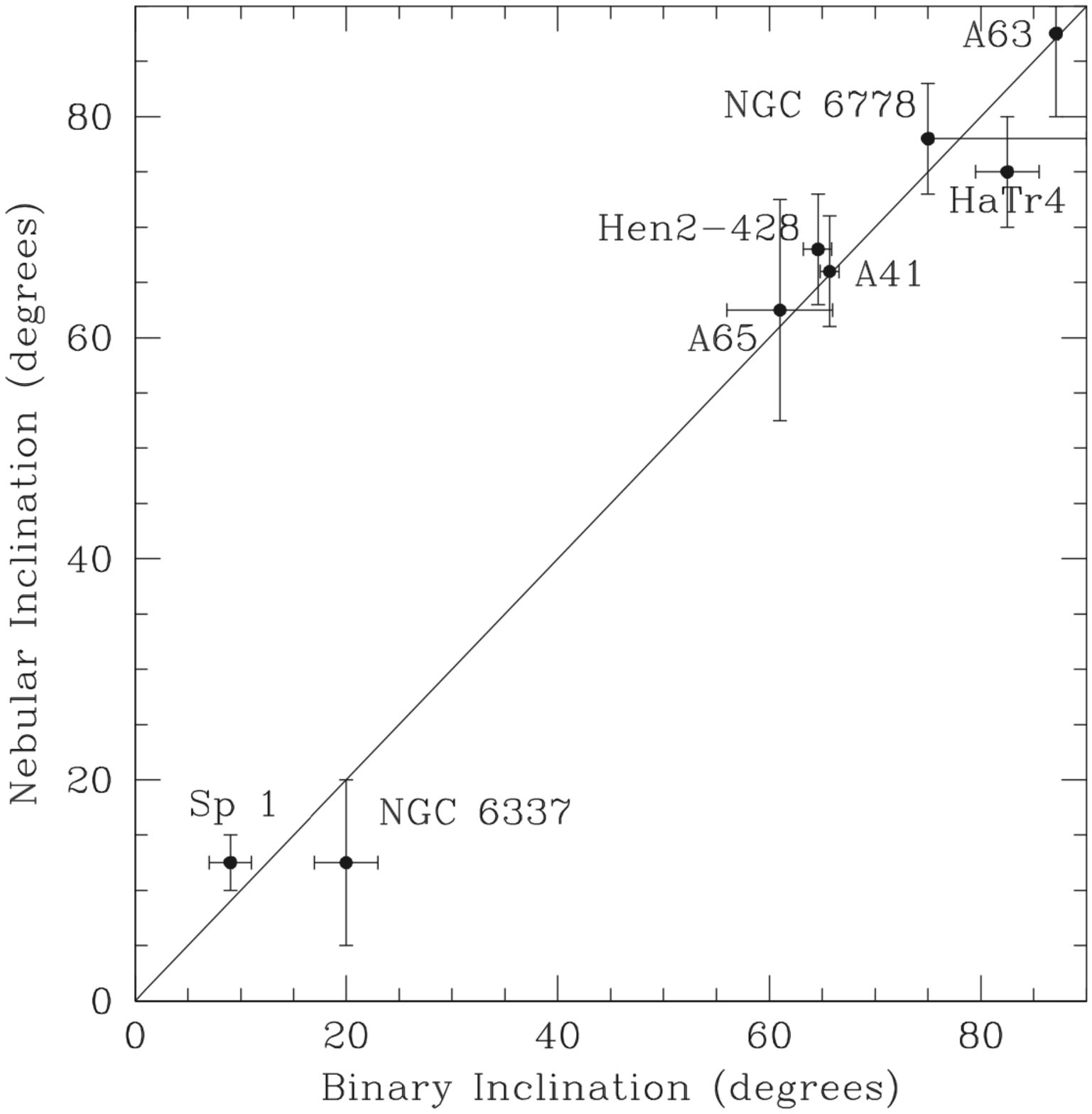} 
\caption{\label{djonesfig2} The observed correlation between binary and nebular inclinations in post-CE PNe, reproduced from \cite{hillwig16}.
}
\end{figure}

A key prediction of all CE models is that, due to the transfer of orbital angular momentum, the envelope should be preferentially ejected in the orbital plane of the binary such that the symmetry axis of the resulting nebula lies perpendicular to this plane \cite{nordhaus06}.  In order to test this prediction, one must determine the morphology and orientation of the PN, via spatio-kinematical modelling \cite{jones12,tyndall12,huckvale13}, as well as the inclination of the central binary, via simultaneous modelling of light and radial velocity curves (see figure \ref{djonesfig1}; \cite{jones15,jones18}).  To date, only eight systems have been the subject of sufficiently detailed study in order to test the prediction, however all eight systems present with the expected correlation (see figure \ref{djonesfig2}). The probability of chance alignment amongst all systems is less that one in a million, as such this constitutes a statistically significant demonstration of the direct link between the central binary and the ejected envelope/nebula \cite{hillwig16}.

Beyond the simple correlation between nebular symmetry axes and binary planes, close-binary PNe reveal much about the CE process.  For example, detailed studies have revealed that main sequence companions are always found to be greatly inflated with respect to isolated stars of the same mass \cite{jones15} - with one exception where the companion is found to present with a relatively normal radius, but at such a short period that it is filling its Roche lobe \cite{jones18}.  The observed inflation is believed to be a result of rapid mass transfer onto the companion knocking the star out of thermal equilibrium.  Further evidence for such mass transfer is found in the central star of the Necklace, where the companion is found to be greatly enhanced in Carbon \cite{miszalski13} - a third dredge-up product which can only have been deposited into the atmosphere of the main sequence star via mass transfer from the primary while it was on the AGB.  In isolation, these results do not say much about when this episode of mass transfer must have occurred, other than that it must have been relatively recently.  However, kinematical studies of several post-CE PNe presenting with extended polar structures, like jets, have shown that these features are kinematically older than the central regions of the nebulae - indeed, in the Necklace, the polar caps are found to be twice as old as the ringed, nebular waist \cite{corradi11}.  Given that the nebular waists are thought to represent the ejected CE, these older polar ejections are likely the product of mass transfer immediately prior to entering into the CE.  Further support for such pre-CE mass transfer is found in the PN Fleming~1, where the precession period of the jets is consistent with a much longer, pre-CE orbital period compared to the 1.2 day, post-CE period now observed \cite{boffin12}.

Given that post-CE central stars have only recently left the CE phase, their stellar and orbital parameters can be used to constrain the efficiency of the process - the holy-grail of close-binary population synthesis.  Of particular interest here are the mass and orbital period distributions, both of which can be altered in more evolved post-CE populations (by mass transfer, magnetic braking, etc.).  The observed orbital period distribution shows a strong peak at periods less than one day, while CE models generally predict a significant number of longer period systems at longer periods \cite{demarco08}.  This makes the recent spate of discoveries of longer period systems \cite{manick15,sowicka17,miszalski18}, all of which would not have been detected via photometric monitoring, particularly intriguing - perhaps indicative of a missing long-period population that has, thus far, evaded detection \cite{demarco04,jones17b}.  Furthermore, in this longer period regime one finds the most massive companion post-CE companion known (not just in terms of central stars of PNe but of all post-CE systems, in general) - that of NGC~2346 weighing in at more than 3.5 M$_\odot$ \cite{brown18} - making this an especially interesting system with which to probe the apparent connection between CE efficiency and binary parameters \cite{davis12}

\section{Double-degenerates and supernovae Ia progenitors}

Based on the shapes of their discovery light curves a significant fraction of post-CE central stars of PNe are thought to be double-degenerate (DD) systems \cite{hillwig10}.  Given the intrinsic difficulty in their detection (DD central stars will not present photometric variability except at very high orbital inclinations and/or very short orbital periods \cite{boffin12,jones17a}), the true fraction of DD central stars is likely much higher.  This has critical implications for our understanding of other astrophysical phenomena - in particular, the cosmologically important supernovae type \textsc{i}a (SNe \textsc{i}a), used to explore the expansion of the Universe at high redshifts (ultimately leading to the award of the 2011 Nobel Prize in Physics).  The merger of two white dwarfs in a close-binary system may represent the main, or even sole, pathway by which SNe \textsc{i}a occur \cite{maoz14}, however to-date no bona-fide progenitor system has been discovered.  Such a system has to meet two main criteria: the total mass of the system must be greater than the Chandrasekhar mass, and the orbital separation should be small enough that the system will merge in less than the age of the Universe.  

Given the apparent over-abundance of DD central stars, as well as the observation that a significant fraction of SNe are found to explode in circumstellar environments consistent with a remnant PN \cite{tsebrenko15}, it is no surprise that the most promising candidate SN \textsc{i}a progenitors have been found inside PNe.  The central star system of TS~01 is a short-period DD system the total mass of which may be greater than the Chandrasekhar mass, however  the uncertainties on the derived total mass encompass sub-Chandrasekhar values \cite{tovmassian10}.  A second candidate is found in the central star of Hen~2-428, where the simultaneous light and radial velocity curve modelling by Santander-Garc\'ia et al.\ derived a total mass of 1.76$\pm$0.26 M$_\odot$ and a time to merger of approximately 700 millions years \cite{santander-garcia15}.  However, the modelled parameters lie far from evolutionary tracks, leading Garc\'ia-Berro et al.\  to question the DD nature of the system, instead proposing that the observed spectral line profiles could arise from a white dwarf and main sequence companion combined with variable nebular line emission \cite{garcia-berro16}.   Recent analyses rule out this hypothesis, clearly demonstrating conclusively that the system comprises two white dwarfs \cite{finch18,reindl18}.  However, Reindl et al.\ find that the measured radial velocity semi-amplitude of the stars (critical in determining their masses) differ depending on the line or lines used - repeating the analysis of Santander-Garc\'ia et al. the results agree to within uncertainties, but including other absorption lines of He \textsc{ii} they derive lower semi-amplitudes and, therefore, a sub-Chandrasekhar total mass \cite{reindl18}.  While there is no clear explanation for a dependence of the radial velocity on the choice of spectral line, the lower masses do bring the derived solution closer to evolutionary tracks.  However, caution must still be exercised in adopting the lower total mass as fact given that, as highlighted by Miller-Bertolami \cite{miller-bertolami17}, these evolutionary tracks do not and, thus far, cannot account for a CE evolution (the exact details of which are far from being understood) and, similarly, it is unclear to what extent the tracks are valid for over-contact systems \cite{reindl18}.  As such, the central stars of TS~01 and Hen~2-428 still represent plausible and, to-date, the strongest candidates to be SN \textsc{i}a progenitor systems.

 \section{Conclusions}
 
 Central star binarity plays a key role in the formation and evolution of PNe, with the clearest impact being found in post-CE systems.  Post-CE PNe are important laboratories for constraining the CE process and, therefore, for understanding the formation of a wide-range of astrophysical phenomena.  The continued presence of the PN ensures that the central star system has only recently exited the CE phase and has not yet had time to adjust, and furthermore as the PN is formed from the remnant envelope it offers a unique opportunity to study the ejection process.  Recent studies of post-CE PNe and their central stars have revealed important information about the CE phase, including evidence of a period of intense mass transfer from the primary onto the companion immediately before entering the CE phase.  Studies of the population as a whole will be key in constraining the CE efficiency and its dependencies - critical in all population synthesis studies, which will have a key role to play in interpreting the results of the next generation of variability surveys including those from the Large Synoptic Survey Telescope (LSST) and the space-based gravitational wave detector the Laser Interferometer Space Antenna (LISA) \cite{belczynski08,korol17}.
 
 Amongst the other phenomena the understanding of which is greatly impacted by studies of post-CE PNe are SNe \textsc{i}a.  Beyond simply constraining the CE, through which all models of SN \textsc{i}a formation pass, PNe have much to reveal about individual models - in particular, for the DD merger scenario.  DD central stars appear over-abundant amongst the total post-CE central star population.  Given that the strongest candidate SN \textsc{i}a progenitor systems are found in PNe, follow-up observations and modelling of other DD central stars may prove a happy hunting ground in the search for a bona-fide progenitor system.  Even if they do not reveal a robust detection of a progenitor, these studies may prove crucial in constraining the importance or even viability of the DD merger scenario, greatly furthering our understanding the origin of SNe \textsc{i}a and what makes them such astonishingly accurate standard candles.


%
%
\small  
%
\section*{Acknowledgments}   
%
This research has been supported by the Spanish Ministry of Economy and Competitiveness (MINECO) under the grant AYA2017-83383-P.
%

%
\end{document}